%% file: main.tex
\documentclass[conference]{IEEEtran}
\IEEEoverridecommandlockouts

\usepackage{cite}
\usepackage{amsmath,amssymb,amsfonts}
\usepackage{algorithm}
\usepackage{algorithmic}
\usepackage{graphicx}
\usepackage{textcomp}
\usepackage{xcolor}
\usepackage{acronym}
\usepackage{url}
\usepackage{subfigure}

\acrodef{nmf}[NMF]{Non-negative Matrix Factorization}
\acrodef{mir}[MIR]{Music Information Retrieval}
\acrodef{stft}[STFT]{Short Time Fourier Transform}

\iffalse

\newcommand\dominique[1]{#1}
\else 

\newcommand\dominique[1]{#1}
\fi

\newcommand{\XX}{\ensuremath{\mathbf{X}}}
\newcommand{\HH}{\ensuremath{\mathbf{H}}}
\newcommand{\NN}{\ensuremath{\mathbf{N}}}
\newcommand{\YY}{\ensuremath{\mathbf{Y}}}
\newcommand{\MAE}{\mathrm{MAE}}
\DeclareMathOperator*{\argmax}{argmax}

\begin{document}

\title{DJ Mix Transcription with Multi-Pass Non-Negative Matrix Factorization}

\author{
    \IEEEauthorblockN{Étienne Paul André}
    \IEEEauthorblockA{
        \textit{UMR STMS -- Ircam, Sorbonne Universités, } \\
        \textit{CNRS, Ministère de la Culture}\\
        Paris, France \\
        andre@ircam.fr
    }
        
    \and
        
    \IEEEauthorblockN{Dominique Fourer}
    \IEEEauthorblockA{
        \textit{IBISC (EA4526)} \\
        \textit{Univ. Evry Paris-Saclay} \\
        \'Evry-Courcouronnes, France \\
        dominique.fourer@univ-evry.fr
    }
        
    \and
        
    \IEEEauthorblockN{Diemo Schwarz}
    \IEEEauthorblockA{
        \textit{UMR STMS -- Ircam, Sorbonne Universités, } \\
        \textit{CNRS, Ministère de la Culture}\\
        Paris, France \\
        schwarz@ircam.fr
    }
}
    
\maketitle

\begin{abstract}
  \input{abstract}
\end{abstract}



  \input{introduction}
  \input{djmixmodel}

 \input{multipassnmf}

  \input{results}

\input{conclusion}
\bibliographystyle{ieeetr}
\bibliography{IEEEabrv, main}
                
\end{document}

%% file: abstract.tex
DJ mix transcription is a crucial step towards DJ mix reverse engineering, which estimates the set of parameters and audio effects applied to a set of existing tracks to produce a performative DJ mix. We introduce a new approach based on a multi-pass NMF algorithm where the dictionary matrix corresponds to a set of spectrogram slices of the source tracks present in the mix.
The multi-pass strategy is motivated by the high computational cost resulting from the use of a large NMF dictionary. The proposed method uses inter-pass filtering to favor temporal continuity and sparseness and is evaluated on a publicly available dataset. 
Our comparative results considering a baseline method based on dynamic time warping (DTW) are promising and pave the way of future NMF-based applications.

%% file: introduction.tex
\section{Introduction}
The transcription of DJ mixes is a challenging problem which aims at estimating how music tracks are combined into a DJ set in time and predicts the digital audio processing parameters (eg. fades, equalization, effects).  It allows further \ac{mir} research into DJ practices for musicology and computer support or automation of DJing.
Mix transcription often requires the following sub-tasks~\cite{SchwarzFourer-lncs2021-dj-mix-reverse-engineering}:
%
\iftrue

{\bf Identification} of the tracks contained in the mix to obtain the playlist, {\bf Alignment} of each track to determine the start and end time in the mix and the speed changes applied by the DJ to achieve beat-synchronicity, {\bf Unmixing} to estimate the cue regions where cross-fading
is applied and the curves for gain and equalization, and also the parameters of other additional effects (dynamic range compression, echo, etc.), {\bf Content and context analysis} to derive the genre, social tags and other metadata of the music that can inform us about the choices a DJ makes when creating a mix.

\else

\begin{LaTeXdescription}
\newcommand{\mitem}[2][ ]{\item[#1] #2}
\mitem[Identification] {of the tracks contained in the mix to obtain the playlist,}
\mitem[Alignment]      {of each track to determine its start and end time in the mix and the speed changes that were applied by the DJ to achieve beat-synchronicity,}
\mitem[Unmixing]       {to estimate the cue regions where cross-fading
is applied and the curves for gain and equalization, and also the parameters of other additional effects (dynamic range compression, echo, etc.),}
\mitem[Content and context analysis] {to derive the genre, social tags and other metadata of the music that can inform us about the choices a DJ makes when creating a mix.}
\end{LaTeXdescription}

\fi

While the identification sub-task is well handled by fingerprinting~\cite{cano2005review, sixOlafLightweightPortable2023, SonnleitnerArztWidmer-ismir2016-landmark, wangIndustrialStrengthAudioSearch}, our work focuses on the alignment and unmixing tasks, and paves the way for future work based on DJ-related MIR content analysis tasks~\cite{lerch2012introduction}.
Thus, we introduce a new method based on multi-pass non-negative matrix factorization (NMF) that is able to extract arbitrary time-warping transformations, including skips or loops, and the mixing gains, while being robust to noise.

\section{Related Work}\label{sec:related}

Several existing works focus on \emph{studio mixes}, where a stereo track is produced from multi-track recordings and software instruments by means of a mixing desk or DAW~\cite{Perez-waspaa2009-automatic, Maddams-dafx2012-autonomous, Mansbridge-aes2012-autonomous, Brecht-ismir2014-analysis-sshort, DeMan-aes2014-open-multitrack, DeMan-innomusic2016-crowd}.
The term \emph{mix reverse engineering} (in the context of multi-track studio mixing) was first used for a method to invert linear processing (gains and delays, including short FIR filters typical for EQ) and some dynamic processing parameters (compression)~\cite{BarchiesiReiss-jaes2010-reverse-engineering}, of interest for our aim of DJ unmixing.
The unmixing problem was already tackled for radio broadcast mixes~\cite{RamonarRichard-dafx2011-fader-estimator}, i.e. retrieving the fader positions of the mixing desk for several known input signals (music tracks, jingles, reports), and one unknown source (the host and guests' microphones in the broadcast studio).
%
These two latter references both assume having sample-aligned source signals at their disposal, with no time-scaling applied, unlike our use-case where each source track only covers part of the mix, can appear partially, repeatedly, and out of order, and can be time-scaled for beat-matched mixing.
More recent work treats alignment~\cite{ewertHighResolutionAudio2009, ramonaAutomaticAlignmentAudio2011, werthen-msc2018-ground-truth-dj-mixes, kimComputationalAnalysisRealWorld2020, SchwarzFourer-lncs2021-dj-mix-reverse-engineering, yangAligningUnsynchronizedPart2021, sixDiscStitchAudiotoaudioAlignment2022}
and mix parameter estimation for unmixing~\cite{werthen-msc2018-ground-truth-dj-mixes, SchwarzFourer-ismir2017lbd-unmixing, kimJointEstimationFader2022}. 
There is rare work related to analysis~\cite{fourer2017a} and inversion of non-linear processing applied to the signal such as dynamic-range compression~\cite{Gorlow-itaslp2013-inversion-compression} which remains challenging and full of interest for unmixing and source separation.
Other works on information retrieval from DJ mixes treats content-based analysis of playlist choices~\cite{Kell-ismir2013-empirical-analysis-track-selection-dance-music}, and track boundaries estimation in mixes~\cite{Scarfe-ijeis2014-Segmentation-Electronic-Dance-Music, Glazyrin14-ismir2014-Towards-Automatic-Content-Based-Separation-of-DJ-Mixes}.

Existing datasets with ground truth DJ mix annotations can be automatically generated~\cite{SchwarzFourer-ismir2018lbd-unmixdb}, hand-crafted~\cite{werthen-msc2018-ground-truth-dj-mixes},
or derived from collaborative annotations~\cite{kimComputationalAnalysisRealWorld2020}.

This paper extends our previous study \cite{SchwarzFourer-lncs2021-dj-mix-reverse-engineering}
replacing multi-scale alignment using dynamic time warping and subsequent estimation of fade curves by a novel method of combined estimation via multi-pass NMF.
It is organized as follows: Section~\ref{sec:model} presents our DJ mix model and our multi-pass NMF method is described in Section~\ref{sec:nmf}. Comparative results are presented in Section~\ref{sec:results} before a conclusion in Section~\ref{sec:conclusion}.

%% file: djmixmodel.tex
\section{Model of the DJ Mixing Process}\label{sec:model}

\subsection{DJ Mixing Model}

\begin{figure}[!ht]
 \subfigure[DJ Mixing process]{\includegraphics[width=0.24\textwidth]{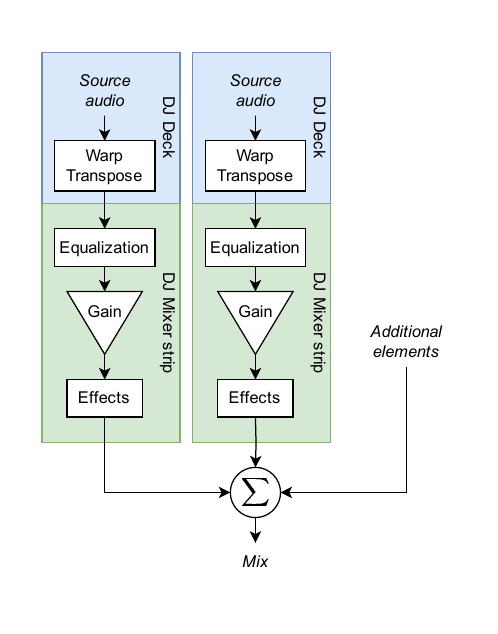}\label{fig:dj-process}}
 \subfigure[Mixing model]{\includegraphics[width=0.24\textwidth]{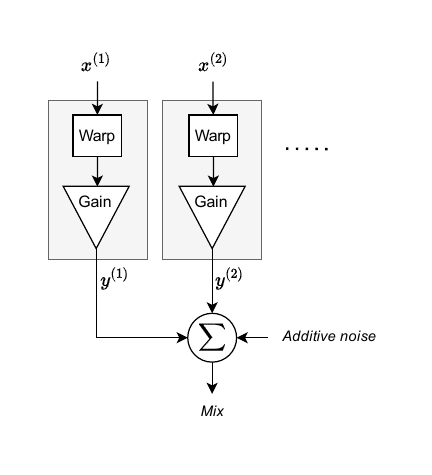}\label{fig:dj-model}}
 \caption{Generalized DJ mixing process and model.}
\end{figure}

Despite the variety of DJ hardware and software options, which come with different features, the DJ's core workflow remains the same across all setups. This consistency allows us to outline a general signal path from the recorded tracks to the final mixed output, as illustrated in Fig. \ref{fig:dj-process}.

The process can be described as follows:
\begin{itemize}
    \item Two or more DJ decks or turntables are used as signal sources. They play pre-recorded tracks and apply pitch-shifting and time-warping.
    \item The signal from the decks is routed to the DJ mixer, which performs a weighted sum of the input signals after an equalization stage. The mixer can also apply various effects (e.g. reverberation, distortion, bit-crushing) and additional elements (e.g. drum machines, speech, sirens).
\end{itemize}

We disregard pitch-shifting and equalization to simplify this model to a simple weighted sum of time-warped tracks, illustrated in Fig.~\ref{fig:dj-model}. 
Effects such as reverberation, distortion, and positive equalization are assumed to be strictly additive and are modeled by additive noise.

\subsection{Capturing Warping and Gain Modulation}
\dominique{Assuming the $I$ source tracks of the mix are known, let \(x^{(i)}\) be the signal of length \(T^{(i)}\) corresponding to the input track \(i \in \{1,2, \ldots, I\}\). The spectrogram matrix \( \XX^{(i)} \in \mathbb{R}^+_{M \times T^{(i)}} \) of \(x^{(i)}\) is defined as the
square modulus of the \ac{stft} computed using an analysis window \(w\).}
%
We model the time-warping and gain-modulation operations by:
\begin{itemize}
    \item
          \(f^{(i)}:\tau \mapsto t\) a time-warping injective function that maps a mix
          time step \(\tau\) to a track time step \(t\).
    \item
          \dominique{\(g^{(i)}\) the positive gain function.} 
\end{itemize}

Let \(y^{(i)}\) be a signal of length \(K\), time-warped by \(f^{(i)}\), and gain-modulated by \(g^{(i)}\), transformation of \(x^{(i)}\). The spectrogram matrix \(\YY^{(i)} \in \mathbb{R}^+_{M \times K}\) of \(y^{(i)}\) is defined as follows, and can be expressed in terms of \(\XX^{(i)}\):

\begin{equation}
    \begin{aligned}
    \YY^{(i)}_{m\tau} & = \left| {g^{(i)}[\tau] \sum_{n = 0}^{M-1} x^{(i)}[n + f^{(i)}[\tau]] w[n] e^{- j2\pi n\frac{m}{M}}} \right|^{2} \\
                & = g^{(i)}[\tau]^{2}\XX^{(i)}_{m,f^{(i)}[\tau]}
    \end{aligned}
\end{equation}

It follows that the warping and gain-modulation operations can be expressed as a matrix product of \(\XX^{(i)}\) with an \emph{activation matrix} \(\HH^{(i)} \in \mathbb{R}^+_{T^{(i)} \times K} \):
\begin{equation}
    \YY^{(i)} = \XX^{(i)}\HH^{(i)}.
    \label{eq:activation}
\end{equation} 
The \emph{ideal activation matrix} \(\tilde{\HH}^{(i)} \), can be expressed
in terms of the Dirac delta function $\delta$ such as:
\begin{equation}
    \tilde{\HH}^{(i)}_{t\tau} = g^{(i)}[\tau]^{2} \delta_{t,f^{(i)}[\tau]}
    \label{eq:ideal}
\end{equation}
%
where the warping and gain-modulation operations can be retrieved from the following estimators:

\begin{equation}
    \tilde{g}^{(i)}[\tau] = \sqrt{\sum_{t = 1}^{T^{(i)}}\tilde{\HH}^{(i)}_{t\tau}}
\end{equation}

\begin{equation}
    \tilde{f}^{(i)}[\tau] = \argmax_{t \in \{1,2,\ldots,T^{(i)}\}}\tilde{\HH}^{(i)}_{t\tau}.
\end{equation}

\subsection{Matrix Representation of the DJ Mixing Process}

Let $\YY \in \mathbb{R}^+_{M \times K}$ be the spectrogram of the mix which
can be expressed as the sum of its transformed source tracks' spectrograms $\YY^{(i)}$, and of the residual noise matrix $\NN$ as:
\dominique{\begin{equation}
    \YY = \NN + \sum_{i = 1}^{I}\YY^{(i)}.
    \label{eq:sum}
\end{equation}
Now, we define two additional matrices \(\bar{\XX}\) and \(\bar{\HH}\) of compatible dimensions so that \(\NN =  \bar{\XX}\bar{\HH}\). We can rewrite Eq.~\eqref{eq:sum} as a product of two large matrices \(\XX \in \mathbb{R}^+_{M \times T} \) and \(\HH \in \mathbb{R}^+_{T \times K}\), where $T = \sum_{i=1}^I T^{(i)}$, by concatenation as:
\begin{equation}\begin{aligned}
    \YY & = \bar{\XX}\bar{\HH} + \sum_{i = 1}^{I}\XX^{(i)}\HH^{(i)} \\
    & = \underset \XX {
            \underbrace{
                \begin{pmatrix}
                    \bar{\XX} ~ \XX^{(1)} ~ \XX^{(2)} ~ \ldots ~ \XX^{(M)}
                \end{pmatrix}
            }
        }
        \underset \HH {
            \underbrace{
                \begin{pmatrix}
                     \bar{\HH} \\ \HH^{(1)} \\ \HH^{(2)} \\ \vdots \\ \HH^{(I)}
                \end{pmatrix}
            }
        }.
    \end{aligned}\end{equation}}

\dominique{Hence, the DJ mix transcription aims to estimate the coefficients of the submatrices \(\HH^{(1)}\) to \(\HH^{(I)}\).} Moreover, an additive noise profile can be learned with the coefficients of \(\bar{\XX}\) and \(\bar{\HH}\). This can be understood as a \ac{nmf} problem which was intensively used for audio source separation tasks \cite{smaragdisNonnegativeMatrixFactor2004}.

%% file: multipassnmf.tex
\section{Multi-pass NMF}\label{sec:nmf}
\subsection{IS-NMF}

\dominique{The NMF algorithm aims to minimise a similarity measure \( \mathcal{D} \) between the \emph{target matrix} \( \YY \) and the \emph{estimated matrix} \( \XX\HH \).
%
To this end, we use the Itakura-Saito divergence defined as:
\begin{equation}
    \begin{split}
        \mathcal{D}_{IS} ( \YY~|~\XX\HH ) = \sum_{f = 1}^{F} \sum_{t = 1}^{T} d_{IS}\left. \left( \YY_{ft}~|~\left. (\XX\HH) \right._{ft} \right) \right. \\
        \text{with } d_{IS}( x~|~y ) = \frac{x}{y} - \ln(xy) - 1
    \end{split}
\label{eq:beta-divergence}
\end{equation}}
for which the estimation of \(\bar{\XX}\) and \(\HH\) can be computed through the 
multiplicative gradient descent algorithm presented in Algorithm~\ref{algo:is-nmf}~\cite{fevotteNonnegativeMatrixFactorization2009}, where \(\odot\) and \( \frac{\cdots}{\cdots} \) respectively stand for the Hadamard (element-wise) product and division. 
\begin{algorithm}
\caption{{IS-NMF} function based on partial multiplicative gradient descent.}
\label{algo:is-nmf}
\begin{algorithmic}
\REQUIRE \(\YY, \XX, \HH\)\\
\dominique{Initialize \HH~randomly}\\
\REPEAT
\STATE \(
    \HH \gets \HH \odot \frac{
        \XX^{T} ( (\XX\HH)^{- 2} \odot \YY )
    }{
        \XX^{T}(\XX\HH)^{- 1}
    }
\)
\STATE \(
    \bar{\XX} \gets \bar{\XX} \odot \frac{
        ( (\XX\HH)^{- 2} \odot \YY ) \bar{\HH}^{T}
    }{
        (\XX\HH)^{- 1}\bar{\HH}^{T}
    }
\)
\UNTIL{convergence criterion is reached}
\RETURN \(\XX, \HH\)
\end{algorithmic}
\end{algorithm}

\subsection{Multi-pass NMF}

Experiments suggest that applying the NMF to music spectrograms produces similarity sub-matrices between \(\YY\) and \(\XX\) instead of the ideal activation matrices described above, on which our estimators yield poor results. 
Using a large overlap (\(> 80~\% \)) for the analysis windows is crucial to emphasize temporal coherence, but not sufficient.

\begin{algorithm}
\caption{Multi-pass NMF}
\label{algo:multipass}
\begin{algorithmic}
    \REQUIRE{\(hlenInit,~hlenTarget,~overlap\)}
    \STATE \(hlen \gets hlenInit \)
    \WHILE{\(hlen > hlenTarget \)}
        \STATE \(wlen \gets hlen \times overlap\)
    
        \FOR{\dominique{\(i = 1\) \TO \(I\)}}
            \STATE \( \XX^{(i)} \gets \text{spectrogram}(x^{(i)}, hlen, wlen)\)
            \IF{\(hlen = hlenInit\)}
                \STATE \(\HH^{(i)} \gets \) random noise
            \ELSE
                \STATE \(\HH^{(i)} \gets \text{interpassFilter}(\HH^{(i)}) \)
                \STATE \(\HH^{(i)} \gets \text{scale}(\HH^{(i)}, hlen) \)
            \ENDIF
        \ENDFOR
        \STATE \(\bar{\XX}, \bar{\HH} \gets\) random noise
        \STATE \(\YY \gets \text{spectrogram}(y, hlen, wlen) \)
        \STATE \(\XX, \HH \gets\) IS-NMF(\YY, \XX, \HH)
        \STATE \(hlen \gets hlen / 2\)
    \ENDWHILE
    \RETURN{\HH}
\end{algorithmic}
\end{algorithm}

\begin{figure}[ht] 
    \centering
    \includegraphics[width=\linewidth]{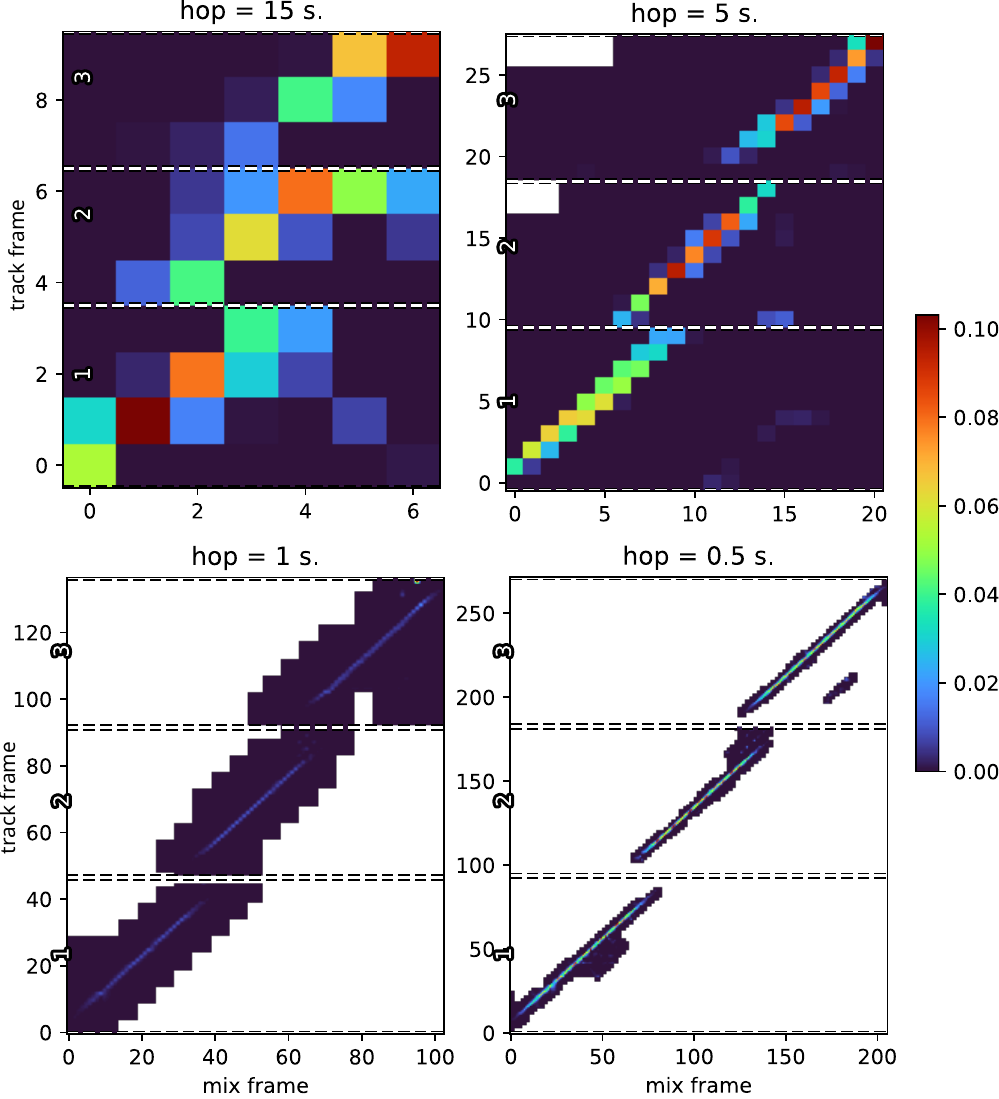}
    \caption{Example of intermediate activation matrices of the multi-pass NMF algorithm for decreasing hop sizes, on a 3-track mix. Zero-valued cells are colored in white. The submatrices' boundaries are indicated by white dashed lines.}
    \label{fig:multipass}
\end{figure}

Thus, we propose a multi-pass NMF algorithm, described in Algorithm \ref{algo:multipass} and illustrated in Fig. \ref{fig:multipass}.
The method involves performing the NMF on spectrograms with increasing temporal resolution. The activation submatrices obtained at each pass are filtered to remove noise, blurred, thresholded then scaled to serve as initialization for the subsequent pass. The iterative process continues until the desired temporal resolution is reached.
Note that,
\newcommand{\myitem}[1][]{(#1)}
    \myitem[1] with each increase in resolution, the activation matrices become less noisy and more sparse,
    \myitem[2] regions of the matrix set to zero by the thresholding step remain zero during the optimization process, thereby avoiding spurious activations,
    \myitem[3] when coupled with an appropriate block-sparse matrix representation, this approach significantly enhances processing efficiency and memory usage, which is efficient
    for processing a large number of tracks in the DJ mix.

\subsection{Inter-pass Filtering Procedure}
The \textit{interpassFilter} procedure of Algorithm 2 performs these steps:
\begin{LaTeXdescription}
\item[Morpohological filtering]
    A line-enhancing filter is applied on each submatrix \(\HH^{(i)}\), inspired by  \cite{mullerEnhancingSimilarityMatrices2006}. The filter kernels \(\mathbf{K}(s,d)\) are one-pixel-wide line kernels of slope \(s\) and length \(d\). A morphological opening (\(\circ\)) is performed and accumulated given slope limits \(s_{min}\) and \(s_{max}\), removing activations shorter than \(d\): 
    \( \forall i : \HH^{(i)} \gets \max_{s \in [s_{min}, s_{max}]} (\HH^{(i)} \circ \mathbf{K}(s,d)) \)
\item[Blurring]
    A Gaussian blur is applied on each submatrix with a small kernel.
\item[Thresholding] Activations smaller than a threshold value, that can be considered negligible, are set to zero.
\end{LaTeXdescription}

%% file: results.tex
\begin{figure*}[!ht]
    \centering
    \includegraphics[width=1\linewidth]{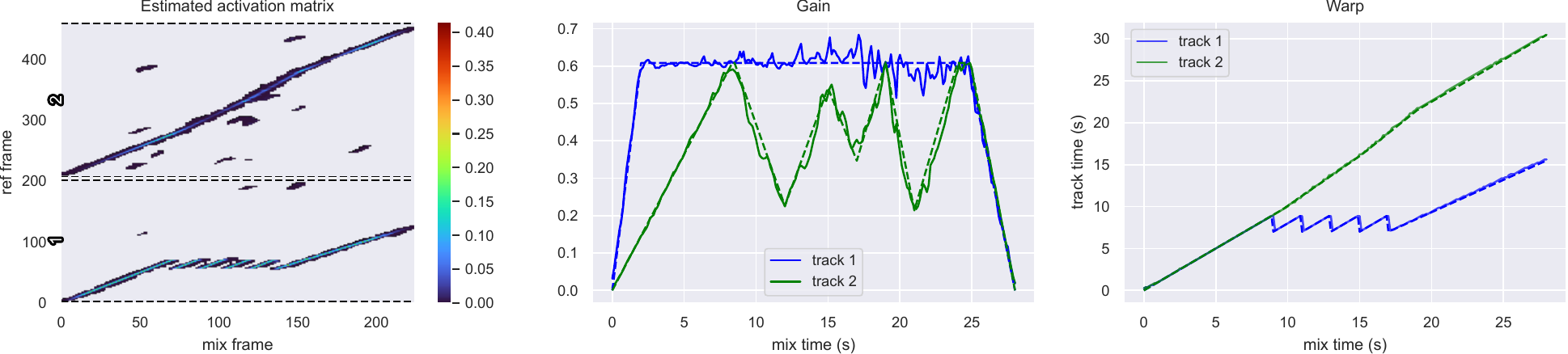}
    \caption{Estimated activation matrix, gain, and warp, on a mash-up of two tracks with complex gain and warp transformations. Ground truth is represented by a dashed line.}
    \label{fig:results}
\end{figure*}

\iffalse

\begin{figure*}[!ht]
    \centering
    \subfigure[]{\includegraphics[width=0.48\textwidth]{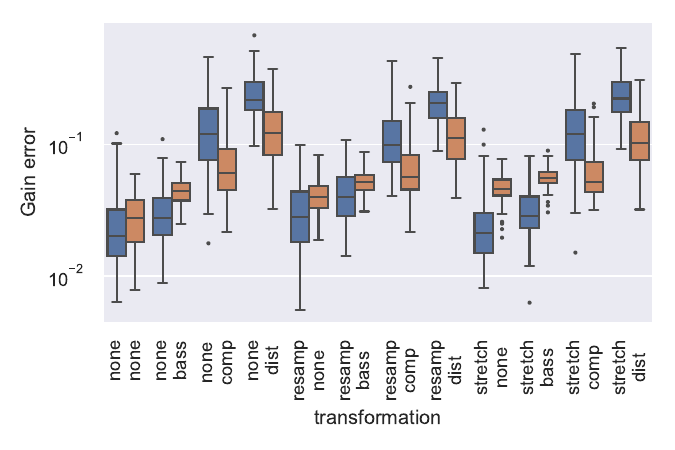}\label{fig:compare-gain}}
    \subfigure[]{\includegraphics[width=0.48\textwidth]{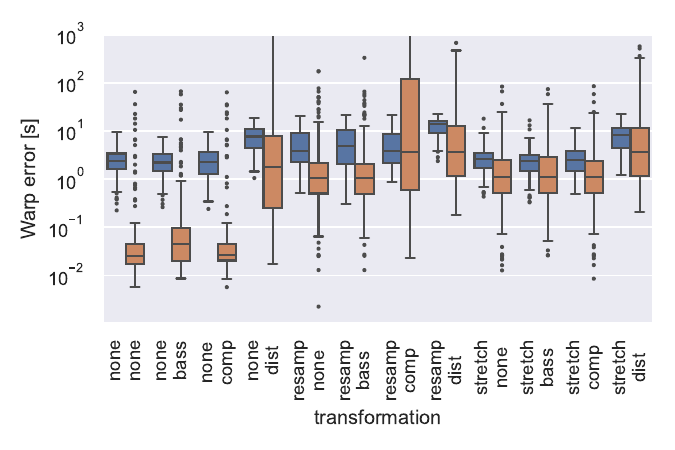}\label{fig:compare-warp}}
    \caption{Box plot of mean absolute estimation error of our method (blue) and DTW-based unmixer (orange), grouped by transformation pair.}
\end{figure*}

\else

\begin{figure*}[!ht]
\includegraphics[width=0.49\textwidth]{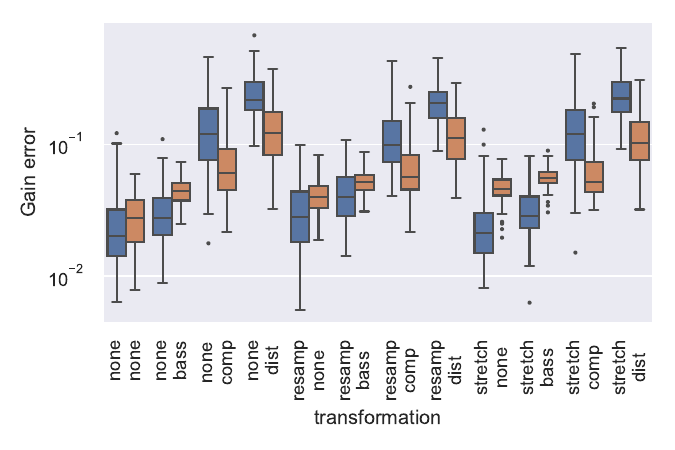}\hfill
\includegraphics[width=0.49\textwidth]{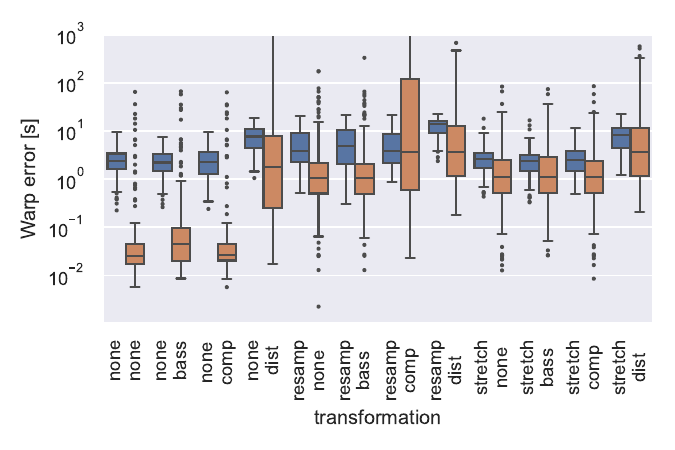}
\caption{Box plot of mean absolute estimation error of our method (blue) and DTW-based unmixer (orange), grouped by transformation pair.}
\label{fig:compare-gain}
\label{fig:compare-warp}
\end{figure*}

\fi

\section{Results}\label{sec:results}
The method was implemented in Python and runs on CPU and/or GPU. The code for reproducing all results is available online.\footnote{\url{https://github.com/etiandre/icassp2025-dj-transcription}}
To illustrate the method's capacity to extract complex warp and gain transformations, Fig. \ref{fig:results} presents the results on a mash-up of two tracks on which fade-in and fade-outs were applied. Track 1 was looped, while track 2 was time-stretched and its gain modulated.
We can see that the activation matrix contains very little errors and thus both the time warping and gain curves $f^{(i)}$ and $g^{(i)}$ are estimated with only small deviations.

The algorithm was evaluated on the \textit{UnmixDB}~\cite{SchwarzFourer-ismir2018lbd-unmixdb} public dataset version 1.1, containing 1931 artificially generated mixes of 3 track excerpts with 3 types of time-warping (\textit{none}, \textit{resampling} changing pitch, \textit{time-stretch} preserving pitch) and~4 effects (\textit{none, bass boost, compression, distortion}).
We compare our method with the existing DTW-based unmixer~\cite{SchwarzFourer-lncs2021-dj-mix-reverse-engineering} in terms of the mean absolute error metric defined by $\MAE(\hat{x},x) = \frac{1}{IK} \sum_{i=1}^I \sum_{\tau=1}^K \left| \hat{x}^{(i)}[\tau] - x^{(i)}[\tau] \right|$, computed for the gain as $g_{error} = \MAE(\tilde{g},g)$ and the warp as $f_{error} = \MAE(\tilde{f},f)$.


We observe in Fig.~\ref{fig:compare-gain}, left, that (1) the gain error is slightly better than the baseline in the no effects and bass boost conditions but slightly worse under the more extreme and unrealistic conditions of compression and distortion, and (2) the model performs well with time-scaling by resampling, even though it is not explicitly pitch-invariant.

However, as shown in Fig.~\ref{fig:compare-warp}, right, our method performs worse in terms of warp error, because it does not impose any specific assumptions on the form of the time-warping function~$f$. In contrast, the baseline method achieves higher accuracy by assuming an affine time-warping function, which aligns well with the dataset's fixed time-stretching factors and avoids complexities such as jumps or irregularities in time-scaling.

%% file: conclusion.tex
\section{Conclusion}\label{sec:conclusion}
We presented a novel application of NMF for the transcription of a DJ mix paving the way for estimating and undoing the articulations and effects a DJ applies to the source tracks.  Our innovative multi-pass algorithm greatly reduces the memory requirements of single-pass NMF by exploiting the inherent sparseness of the activation matrices, making it applicable to realistic mixes of one hour length, and is able to capture loops and jumps.
The results on complex test mixes are promising, although the quantitative evaluation on the \textit{UnmixDB} dataset does not always surpass the state of the art, due to the dataset's unrealistically extreme transformations, and its limitation to constant time-scaling.  
Future more realistic datasets~\cite{Andre-msc2024-dj-mix-transcription} will be able to better show the strengths and limitations of our algorithm; and research into pitch-invariance or estimation, equalization estimation, and regularization of the NMF algorithm could allow to further improve the method.